%%%%%%%%%%%%%%%%%%%%%%%%%%%%%%%%%%%%%%%%%%%%%%%%%%%%%%%%%%%%%%%%%%%%%%%%
% paper "On the diffeomorphism commutators of lattice quantum gravity",% 
% by R. Loll, begins here (tex-file)                                   %
%%%%%%%%%%%%%%%%% fonts, definitions,etc.%%%%%%%%%%%%%%%%%%%%%%%%%%%%%%% 

\input epsf.tex

\font\rmu=cmr10 scaled\magstephalf
\font\bfu=cmbx10 scaled\magstephalf

\font\it=cmti10 scaled \magstephalf

\rmu

\font\rmus=cmr8
\font\rmuss=cmr6
\font\mait=cmmi10 scaled\magstephalf
\font\maits=cmmi7 scaled\magstephalf
\font\maitss=cmmi7
\font\msyb=cmsy10 scaled\magstephalf
\font\msybs=cmsy8 scaled\magstephalf
\font\msybss=cmsy7
\font\bfus=cmbx7 scaled\magstephalf
\font\bfuss=cmbx7
\font\cmeq=cmex10 scaled\magstephalf

\textfont0=\rmu
\scriptfont0=\rmus
\scriptscriptfont0=\rmuss

\textfont1=\mait
\scriptfont1=\maits
\scriptscriptfont1=\maitss

\textfont2=\msyb
\scriptfont2=\msybs
\scriptscriptfont2=\msybss

\textfont3=\cmeq
\scriptfont3=\cmeq
\scriptscriptfont3=\cmeq

\newfam\bmufam  \textfont\bmufam=\bfu
      \scriptfont\bmufam=\bfus \scriptscriptfont\bmufam=\bfuss

\hsize=15.5cm
\vsize=21cm
\baselineskip=16pt   % Double spacing
\parskip=12pt plus  2pt minus 2pt

\def\d{\delta}
\def\e{\epsilon}

\def\semi{\bigcirc\kern-1em{s}\;}

\def\R{{\rm I\!R}}

\def\one{{\mathchoice {\rm 1\mskip-4mu l} {\rm 1\mskip-4mu l}
{\rm 1\mskip-4.5mu l} {\rm 1\mskip-5mu l}}}
\def\Q{{\mathchoice
{\setbox0=\hbox{$\displaystyle\rm Q$}\hbox{\raise 0.15\ht0\hbox to0pt
{\kern0.4\wd0\vrule height0.8\ht0\hss}\box0}}
{\setbox0=\hbox{$\textstyle\rm Q$}\hbox{\raise 0.15\ht0\hbox to0pt
{\kern0.4\wd0\vrule height0.8\ht0\hss}\box0}}
{\setbox0=\hbox{$\scriptstyle\rm Q$}\hbox{\raise 0.15\ht0\hbox to0pt
{\kern0.4\wd0\vrule height0.7\ht0\hss}\box0}}
{\setbox0=\hbox{$\scriptscriptstyle\rm Q$}\hbox{\raise 0.15\ht0\hbox to0pt
{\kern0.4\wd0\vrule height0.7\ht0\hss}\box0}}}}
\def\C{{\mathchoice
{\setbox0=\hbox{$\displaystyle\rm C$}\hbox{\hbox to0pt
{\kern0.4\wd0\vrule height0.9\ht0\hss}\box0}}
{\setbox0=\hbox{$\textstyle\rm C$}\hbox{\hbox to0pt
{\kern0.4\wd0\vrule height0.9\ht0\hss}\box0}}
{\setbox0=\hbox{$\scriptstyle\rm C$}\hbox{\hbox to0pt
{\kern0.4\wd0\vrule height0.9\ht0\hss}\box0}}
{\setbox0=\hbox{$\scriptscriptstyle\rm C$}\hbox{\hbox to0pt
{\kern0.4\wd0\vrule height0.9\ht0\hss}\box0}}}}

\font\fivesans=cmss10 at 4.61pt
\font\sevensans=cmss10 at 6.81pt
\font\tensans=cmss10
\newfam\sansfam
\textfont\sansfam=\tensans\scriptfont\sansfam=\sevensans\scriptscriptfont
\sansfam=\fivesans
\def\sans{\fam\sansfam\tensans}
\def\Z{{\mathchoice
{\hbox{$\sans\textstyle Z\kern-0.4em Z$}}
{\hbox{$\sans\textstyle Z\kern-0.4em Z$}}
{\hbox{$\sans\scriptstyle Z\kern-0.3em Z$}}
{\hbox{$\sans\scriptscriptstyle Z\kern-0.2em Z$}}}}

\newcount\foot
\foot=1
\def\note#1{\footnote{${}^{\number\foot}$}{\ftn #1}\advance\foot by 1}

\def\frac#1#2{{#1\over #2}}
\def\text#1{\quad{\hbox{#1}}\quad}

\font\ch=cmbx12 scaled\magstephalf
\font\ftn=cmr8 scaled\magstephalf

\font\it=cmti10 scaled\magstephalf

\font\titch=cmbx12 scaled\magstep2
\font\titname=cmr10 scaled\magstep2
\font\titit=cmti10 scaled\magstep1
\font\titbf=cmbx10 scaled\magstep2

\nopagenumbers

%%%%%%%%%%%%%%%%%%% title page %%%%%%%%%%%%%%%%%%%%%%%%%%%%%%%%%%
\line{\hfil AEI-039}
\line{\hfil August 12, 1997}
\vskip1cm
\centerline{\titch ON THE DIFFEOMORPHISM-COMMUTATORS OF}
\vskip.5cm
\centerline{\titch LATTICE QUANTUM GRAVITY}
\vskip3.7cm
\centerline{\titname R. Loll}
\vskip.5cm
\centerline{\titit Max-Planck-Institut f\"ur Gravitationsphysik}
\vskip.2cm
\centerline{\titit Schlaatzweg 1}
\vskip.2cm
\centerline{\titit D-14473 Potsdam, Germany}

\vskip4.5cm
\centerline{\titbf Abstract}
\vskip0.2cm
We show that the algebra of discretized spatial diffeomorphism
constraints in Hamiltonian lattice quantum gravity closes without anomalies
in the limit of small lattice spacing. The result holds for arbitrary
factor-ordering and for a variety of different discretizations of
the continuum constraints, and thus generalizes an earlier calculation by
Renteln.

\vfill\eject
\footline={\hss\tenrm\folio\hss}
\pageno=1
%%%%%%%%%%%%%%%%%%%%%%%%%%%%%%%%%%%%%%%%%%%%%%%%%%%%%%%%%%%%%%%%%%%%

\line{\ch 0 Motivation\hfil}

For the most part, our understanding of quantum gravity in four 
dimensions is not sufficiently developed to provide unambiguous tests
for the consistency and appropriateness of any candidate theory.
In the case of usual gauge field theories,
there are well-known criteria to ensure their consistency at the
quantum level, for example, the validity of the Slavnov-Taylor identities.
However, the Fock space methods on which they are based are only of
limited use for gravitational theories in more than two space-time
dimensions. 

Attempts to construct non-perturbative path integral formulations 
of 4-d gravity have so far left the question of their continuation to
Lorentzian signature unanswered. Even disregarding this problem,
it is difficult to get a mathematical handle on the structure of 
the path integral measure.
Canonical approaches circumvent the signature problem and 
allow one to pose (part of) the quantum consistency problem in a 
seemingly clear-cut form: is there an anomaly in the Dirac 
algebra of the Hamiltonian and diffeomorphism constraints or isn't 
there? In the presence of anomalous terms, the quantum Dirac
conditions overconstrain the space of physical states, which 
is physically unacceptable.  
One difficulty of the canonical formulation 
lies in finding appropriate regularized
versions of the constraint operators, without which the question is
known to be ill-posed [1]. A potential drawback of the Hamiltonian 
approaches is the fact that efficient computational techniques for a
numerical study of their quantum properties have yet to be developed.

Schemes that proceed by a direct discretization of space-time or of the
space of all field configurations in order to achieve a regularization 
tend to break the diffeomorphism invariance present in the continuum theory.  
This however is not a fundamental objection, if one considers
the smooth structure of the classical 
theory only as a semi-classical property, which does not 
continue to hold down to smallest length scales. Still it leaves one
with the technical inconvenience of having to work with
an ``approximately diffeomorphism-invariant'' regularized theory. 

Considerable efforts have gone into trying to construct
a canonical quantization of 3+1 gravity in terms of connection
variables, without resorting to a discretization, and with the full 
spatial diffeomorphism group still acting on a suitably defined space 
of quantum states [2-5]. 
However, this approach seems to be running into some difficulties,
among which are a ``non-interacting'' property of typical realizations 
of the quantum 
Hamiltonian, which leads, for example, to the existence of 
unexpected local quantum observables [6,7], vanishing
commutation relations between two such Hamiltonians [8] 
(even in situations where the diffeomorphism group acts non-trivially), 
and anomalous commutation relations between geometric operators that 
classically commute [9,10]. 

This provides an additional incentive for studying a truly discretized
version of this formulation, which does not share any of these
features. On the other hand, like most discretized models, 
it does not carry any obvious, non-trivial 
representation of the classical invariance group of general relativity.
The diffeomorphism group is only recovered in the limit as the
regulator is taken to zero.

In the present piece of work, we will generalize a
result obtained earlier by Renteln [11] in the framework of a 
lattice discretization of 3+1 gravity in a 
connection formulation [11-17]. Various aspects of this version of
lattice quantum gravity have changed since these earlier investigations,
as a result of new developments both in
the continuum loop representations of quantum gravity mentioned above, and
in lattice gravity proper. The kinematic resemblance
with Hamiltonian lattice gauge theory has become even closer since
the theory was rephrased in terms of real $su(2)$-valued connections 
[18], instead of the original $sl(2,\C)$-valued Ashtekar potentials 
[19], in order to 
avoid complications related to the reality conditions that appear in the
complex case. As a by-product, there is now a well-defined scalar
product and a complete set of square-integrable states for each 
finite-size lattice. Although this may not give rise to a scalar product
on the sector of physical states (satisfying all of the constraints), it is
nevertheless remarkable that it should exist at all, given that
inner products are usually hard to come by in quantum gravity. 

We will show that {\it independent of the factor-ordering} and for a
variety of operator symmetrizations,
there are no anomalous terms in the commutator 
algebra of the lattice analogues of the spatial diffeomorphism generators
in the limit of vanishing lattice spacing. 
The computations are already substantial for this subalgebra
of the entire quantum constraint algebra. We will also spell out 
some details of the calculations, that may be of interest in 
attempts to compute the commutator in alternative 
regularization schemes. This result shows that no inconsistencies arise in
the lattice discretization at this level, and paves the way for the
calculation of commutators involving also the Hamiltonian constraint.

\vskip2cm

\line{\ch 1 Introduction\hfil}

A prominent feature of the lattice discretization we are considering
is its resemblance
with canonical lattice gauge theory. Like Hamiltonian gauge
field theory, 3+1 continuum gravity can be formulated in terms of Yang-Mills 
conjugate variable pairs $(A,E)$ of ``Ashtekar-type", with Poisson brackets
$\{A_a^i(x),E_j^b(y)\}=\delta^i_j \delta_a^b \delta^3 (x, y)$.
The variable $A_a^i$ is a spatial $su(2,\R)$-valued gauge potential.

Up to terms proportional to the Gauss law constraint, the Hamiltonian 
constraint in this formulation is the Barbero Hamiltonian [18] (rescaled by  
a factor $({\det E})^{-\frac{1}{2}}$ to make it a density of weight one),
plus -- for the sake of generality  -- a cosmological constant term,

$$
\eqalign{
{\cal H}(x)=&\frac{1}{\sqrt{\det E}}\epsilon^{ijk} E_i^a E_j^b F_{ab}^k -
 \frac{1}{G}(\det E)^{-5/2} \eta_{a_1 a_3 a_4}\eta_{b_1 b_3 b_4}
(E^{a_3}_k E^{a_4}_l E^{b_3}_m E^{b_4}_n\cr
&- 2 E^{a_3}_m E^{a_4}_n E^{b_3}_k E^{b_4}_l) 
E^{a_2}_k E^{b_2}_m 
(\nabla_{a_2} E^{a_1}_l)(\nabla_{b_2} E^{b_1}_n)+
\frac{\lambda}{G} \sqrt{\det E}=0.}\eqno(1.1)
$$

We have chosen the dimensions of the basic variables and 
constants to be $[A]=L^{-3}$, $[E]=L^{0}$, $[G]=L^{2}$, $[\lambda]=L^{-2}$,
where $G$ is Newton's constant. 
Also recall that $E$ transforms as a one-density under spatial 
diffeomorphisms.
In (1.1), $\cal H$ has already been brought into the form of a polynomial
modulo powers of $\sqrt{\det E}$, in order to make its discretization
and quantizion straightforward. The spatial diffeomorphism
constraints are given by

$$
{\cal V}_{b}(x)=E^{a}_{i}(x)F_{ab}^{i}(x)=0.\eqno(1.2)
$$

\noindent Smearing out these four constraints with the lapse and shift 
functions $N(x)$ and $N^{a}(x)$, one arrives at the expressions

$$
\eqalign{ 
& H[N]:=\int d^{3}x\;N(x){\cal H}(x)\cr
& V[N^{a}]:=\int d^{3}x\;N^{b}(x){\cal V}_{b}(x),}\eqno(1.3)
$$

\noindent which satisfy the usual Dirac Poisson algebra. In particular,
for the spatial diffeomorphism generators one derives 

$$
\eqalign{
\{  V[N^{a}],  V[M^{a}] \}=
&\int d^{3}x\,(N^{b}\partial_{b}M^{a}-M^{b}\partial_{b}N^{a})F_{ac}^{i}
E^{c}_{i}\; -\int d^{3}x\,F_{ab}^{i}N^{a}M^{b}(\nabla_{c}E^{c}_{i})\cr
\equiv \, & V[L_{N}M^{a}]- G[ F_{ab}^{i}N^{a}M^{b} ].}
\eqno(1.4)
$$

\noindent which coincides with the diffeomorphism Lie algebra on the 
subspace of phase
space given by the vanishing of the integrated Gauss law constraints
$ G[\Lambda^{i} ]=\int d^{3}x\,\Lambda^{i}(\nabla_{a}E^{a}_{i})$. One
may also redefine the generators $V[N^{a}]$ by adding a suitable term 
proportional to the Gauss law constraint in order to get rid of the extra 
phase-space dependent term on the right-hand side of (1.4) (see, for 
example, [20]).

\vskip2cm

\line{\ch 2 Discretization\hfil}

Before setting up the discretization of operators relevant for
lattice gravity, we start with a brief summary of the basic ingredients 
of Hamiltonian lattice gauge theory [21]. 
For computational simplicity, we take the lattice $\Lambda$ to be a cubic 
$N^3$-lattice with periodic boundary conditions. The cubic symmetry is
convenient but not necessary, as we know from the continuum loop 
quantizations which use imbedded lattices.

The basic quantum operators
associated with each lattice link $l$ are a group-valued $SU(2)$-link
holonomy $\hat U$ (represented by multiplication by $U$),
together with its inverse $\hat U^{-1}$, and a pair
of canonical momentum operators $\hat p^+_i$ and $\hat p^-_i$, where
$i$ is an adjoint index. The operator $\hat p^+_i(n,\hat a)$ is based
at the vertex $n$, and is associated with the link $l$ oriented in the
positive $\hat a$-direction. By contrast, $\hat p^-_i(n+\hat 1_{\hat
a},\hat a)$ is based at the vertex displaced by one lattice unit in
the $\hat a$-direction, and associated with the inverse link
$l^{-1}(\hat a)=l(-\hat a)$. In mathematical terms, the momenta 
$\hat p^+$ and $\hat p^-$ correspond to the left- and right-invariant
vector fields on the group manifold associated with a given link.
The wave functions are elements of
$\times_l L^2(SU(2),dg)$, with the product taken over all links, and
the canonical Haar measure $dg$ on each copy of the group $SU(2)$. 
The basic commutators are

$$
\eqalign{
&[\hat U_A{}^B(n,\hat a),\hat U_C{}^D(m,\hat b)]=0,\cr
&[\hat  p^+_i(n,\hat a),\hat U_A{}^C(m,\hat b)]=
-\frac{i}{2}\,\d_{nm}\d_{\hat a\hat b}\, \tau_{iA}{}^B\hat U_B{}^C
(n,\hat a),\cr
&[\hat  p^-_i(n,\hat a),\hat U_A{}^C(m,\hat b)]=
-\frac{i}{2}\,\d_{n,m+1}\d_{\hat a\hat b}\,\hat U_A{}^B
(n,\hat a) \tau_{iB}{}^C,\cr
&[\hat p^\pm_i(n,\hat a),\hat p^\pm_j(m,\hat b)]=
\pm i\, \d_{nm}\d_{\hat a\hat b}\, \e_{ijk}\, \hat p_k^\pm
(n,\hat a),\cr
&[\hat p^+_i(n,\hat a),\hat p^-_j(m,\hat b)]=0,}\eqno(2.1)
$$

\noindent where $\epsilon_{ijk}$ are the structure constants of $SU(2)$.
(There are analogous Poisson bracket relations for the corresponding
classical lattice variables.)
The commutation relations for the inverse holonomy operators can be easily
deduced from (2.1).  
The normalization for the $SU(2)$-generators $\tau_i$ is such that
$[\tau_i,\tau_j]=2\,\epsilon_{ijk}\tau_k$, ${\rm 
Tr}\,\tau_{i}\tau_{j}=-2\delta_{ij}$ and $A_a=A_a^i \tau_i/2$.

In order to relate discrete lattice expressions with their continuum
counterparts, one uses power series expansions in the so-called
lattice spacing $a$, which is an unphysical parameter with dimension
of length. For the basic classical lattice variables, these are

$$
\eqalign{
U_A{}^B(\hat b)&=1_A{}^B+a\,G\, A_{bA}{}^B+ \frac{a^2 G}{2}
(\partial_b A_b+GA_b^2)_A{}^B+O(a^3),\cr
p_i^\pm (\hat b)&=a^2\,G^{-1}\, E^b_i \pm a^3\, G^{-1}\,
\nabla_b E_i^b+ O(a^4).}\eqno(2.2)
$$

We will assume that similar expansions continue to be valid in 
the quantum theory. Note that Newton's constant $G$ appears in 
(2.2) since the dimensions of the
basic gravitational variables $A$ and $E$ differ from those of the
corresponding Yang-Mills phase space variables. 

Using the expansions (2.2), one obtains unambiguous continuum 
limits of composite
classical lattice expressions by extracting the coefficient of the
lowest-order term in the $a$-expansion. The converse is not true:
there is no unique lattice discretization of a continuum expression,
since one may always add to the lattice version terms of higher
order in $a$ which do not contribute in the continuum limit. We will
consider in the following different symmetrizations for the
lattice diffeomorphism generators.

\vskip2cm

\line{\ch 3 Further preparations\hfil}

As a first step in the commutator calculation of two lattice
diffeomorphisms, one needs to expand the classical link holonomies in
a neighbourhood of the lattice vertex $n_{0}$ at which the local commutator
will be computed. For simplicity we choose a local lattice coordinate
system such that the vertex $n_{0}$ coincides with its origin, that is,
$n_{0}=(0,0,0)$. There are six nearest neighbours, which are one 
lattice unit away, for example, the vertices $n=(\pm 1,0,0)$ in 
positive and negative $\hat 1$-direction. Because of the non-locality 
of the diffeomorphism generators, all relevant link holonomies lie on
a cube of edge length 2 about the origin.

Using the defining formula for the holonomy along a parametrized 
path $\gamma$  
between two points $x_{1}$ and $x_{2}$, with $\gamma (0)=x_{1}$ and
$\gamma (1)=x_{2}$,

$$
U(x_{1},x_{2})=\one +\sum_{n=1}^{\infty}\int_{0}^{1} ds_{1} 
\int_{s_{1}}^{1} ds_{2}\ldots \int_{s_{n-1}}^{1} ds_{n}\;
A(\gamma(s_{1})) A(\gamma(s_{2})) \ldots A(\gamma(s_{n})),
\eqno(3.1)
$$

\noindent one expands this into a power series in the lattice spacing 
$a$ for the special choice $x_{1}=(0,0,0)$, $x_{2}=(a,0,0)$, say. We 
will need terms up to power $a^{3}$ for our computation. For the 
holonomy in positive $\hat b$-direction based at $(0,0,0)$, one
obtains thus

$$
U((0,0,0),\hat b)=\one+aA_{b}+\frac{a^{2}}{2}(\partial_{b}A_{b}+A_{b}^{2})+
\frac{a^{3}}{3!}(\partial_{b}^{2}A_{b}+(\partial_{b}A_{b}) A_{b} +
2A_{b}(\partial_{b}A_{b})+A_{b}^{3})+ O(a^{4}),\eqno(3.2)
$$

\noindent where we now have set $G=1$. 
(Note that at order $a^{3}$ this differs from the expression given in [11].) 
The other positively oriented link holonomies are
computed similarly, as Taylor expansions with respect to the
local cartesian coordinate system spanned by the lattice. From those,
the inverse holonomies (associated with the opposite link orientation)
are computed order by order using $UU^{-1}=\one$. Expansions for the
link momenta are obtained in an analogous manner, using (2.2). 

Two useful checks for possible errors are given by calculating
holonomies $U_{\sqcap \mskip-12mu \sqcup}$ around plaquettes, i.e.
shortest closed lattice paths of edge length four. Denoting by
$U_{\sqcap \mskip-12mu \sqcup_{12}}$ the plaquette that starts with
a link in positive $\hat 1$-direction, followed by links in
directions $\hat 2$, $-\hat 1$ and $-\hat 2$, one has the expansions

$$
{\rm Tr}\, U_{\sqcap \mskip-12mu \sqcup_{12}}=2+O(a^{4}),\qquad
{\rm Tr}\, (U_{\sqcap \mskip-11.5mu \sqcup_{12}}\tau_{i})=
a^{2}F_{12}^{i}+O(a^{3}).\eqno(3.3)
$$

A natural discretization for the smeared spatial diffeomorphism
generators, (second line of (1.3)), is schematically given by

$$
\sum_{n} N^{\rm latt}(n,\hat a) \,{\rm 
Tr}\,(U_{\sqcap \mskip-11.5mu \sqcup_{ab}}\tau_{i})p_{i}(n,\hat b)=:
\sum_{n} {\cal V}^{\rm latt}[N,n) .\eqno(3.4)
$$ 

\noindent where we have not yet specified the details of the implicit
summation over the indices $\hat a$ and $\hat b$.
We will consider four distinct ways of symmetrizing the local lattice
expressions 
$N^{\rm latt}(n,\hat a) \,{\rm Tr}\,( U_{\sqcap \mskip-11.5mu \sqcup_{ab}}
\tau_{i})p_{i}(n,\hat b)$.
Let us introduce the notation $N^{\pm}(n,\hat a)$ for the lattice
shift functions at vertex $n$, associated with the link in positive
and negative $\hat a$-direction respectively. This notation conforms 
with the one for the momenta $p^{\pm}(n,\hat a)$. In order for (3.4) 
to have the correct continuum limit as $a\rightarrow 0$, we require 
that

$$
\eqalign{
&N^{+}(n,\hat b)\rightarrow \frac{1}{a}N^{b}(x)+\tilde{N}^{b}(x)+O(a^{1})\cr
&N^{-}(n,\hat b)\rightarrow \frac{1}{a}N^{b}(x)+\tilde{\tilde{N}}{}^{b}(x)
+O(a^{1}).} \eqno(3.5)
$$

\noindent We leave the zeroth-order functions $\tilde{N}^{b}(x)$ and
$\tilde{\tilde{N}}{}^{b}(x)$ unspecified; the final result will not depend
on this choice. For the local continuum expression of the form 
$N^{2}F_{23}^{i}E_{i}^{3}$, say, we will consider four different
lattice representations:
\item{(a)} no symmetrization:
$$
N^{+}(n,\hat 2) {\rm Tr}\, (U_{\sqcap \mskip-12mu \sqcup_{23}}) 
p_{i}^{+}(n,\hat 3);
$$
\item{(b)} symmetrization over shift functions:
$$
\frac{1}{2} ( N^{+}(n,\hat 2) {\rm Tr}\, (U_{\sqcap \mskip-12mu \sqcup_{23}}) 
p_{i}^{+}(n,\hat 3) +
N^{-}(n,\hat 2) {\rm Tr}\, (U_{\sqcap \mskip-12mu \sqcup_{3,-2}}) 
p_{i}^{+}(n,\hat 3));
$$
\item{(c)} symmetrization over momenta:
$$
\frac{1}{2}(N^{+}(n,\hat 2) {\rm Tr}\, (U_{\sqcap \mskip-12mu \sqcup_{23}}) 
p_{i}^{+}(n,\hat 3) +
N^{+}(n,\hat 2) {\rm Tr}\, (U_{\sqcap \mskip-12mu \sqcup_{-3,2}}) 
p_{i}^{-}(n,\hat 3) );
$$
\item{(d)} symmetrization over both shift functions and momenta:
$$
\eqalign{
\frac{1}{4} (& N^{+}(n,\hat 2) {\rm Tr}\, (U_{\sqcap \mskip-12mu \sqcup_{23}}) 
p_{i}^{+}(n,\hat 3) +
N^{-}(n,\hat 2) {\rm Tr}\, (U_{\sqcap \mskip-12mu \sqcup_{3,-2}}) 
p_{i}^{+}(n,\hat 3) +\cr
&N^{+}(n,\hat 2) {\rm Tr}\, (U_{\sqcap \mskip-12mu \sqcup_{-3,2}}) 
p_{i}^{-}(n,\hat 3) +
N^{-}(n,\hat 2) {\rm Tr}\, (U_{\sqcap \mskip-12mu \sqcup_{-2,-3}}) 
p_{i}^{-}(n,\hat 3) );}
$$ 

Note that in each case we sum over terms that are maximally localized 
on the lattice, i.e. consist only of holonomies and momenta located
on a single lattice plaquette. Choice (d) was the one considered in [11].

\vskip2cm

\line{\ch 4 The commutator\hfil}

We now come to the actual computation of the commutator of two
lattice-regularized diffeomorphism constraints. As is well known,
the algebra of the discretized constraints (3.4) does not close, not even
at the level of the classical Poisson algebra, because the result is
a sum of terms, where each term may extend over up to two 
plaquettes. That is, the Poisson commutator is not a linear combination
of terms of the form (3.4). 
However, if the discretization is consistent, the commutator should
yield the continuum result to lowest order in the lattice spacing
$a$ as $a\rightarrow 0$. 

We will perform the corresponding quantum computation, and for 
definiteness will choose initially the factor-ordering for the
${\rm Tr}\,(\hat U_{\sqcap \mskip-11.5mu \sqcup_{ab}}
\tau_{i})\hat p_{i}(n,\hat b)$-terms with all 
momenta $\hat p$ to the right.  

To obtain the contribution at a given lattice vertex $n_{0}$ to the
commutator of two smeared lattice diffeomorphisms 
$[ \sum_{n} \hat{{\cal V}}[N,n), \sum_{m} \hat{{\cal V}}[M,m)]$,
one has to collect all terms where a momentum based at $n_{0}$ 
acts on either a holonomy or another momentum based on the same link.
One quickly realizes that there are not only contributions from
terms in $[ \hat{{\cal V}}[N,n_{0}), \hat{{\cal V}}[M,n_{0})]$,
but also contributions from vertices close-by, either from nearest
neighbours (e.g. $n_{0}\pm (1,0,0)$ -- there are 6 such vertices) or 
from vertices across a plaquette diagonal (e.g. $n_{0}\pm (1,1,0)$ -- 
there are 12 such vertices). We will see below that these 
non-local contributions are indeed necessary for obtaining the
correct result. 

Without loss of generality, we will consider local lattice smearing functions
$N(n_{0},\hat a)=(N^{1},0,0)$ and $M(n_{0},\hat a)=(0,M^{2},0)$.
For the fully symmetrized version of the local constraints
$\hat{\cal V}[N,n)$, there are 36 non-vanishing commutators between terms 
based at $n_{0}$ and 72 non-vanishing commutators involving also terms
based at neighbouring vertices. The commutators are evaluated using
the basic commutation relations (2.1) and various epsilon-function identities.
The most time-consuming task is the expansion and subsequent simplification
of the resulting expressions in powers of $a$. It turns out that the 
holonomies have to be expanded to third order, i.e. up to terms cubic in 
the local connection form $A(x)$, and there are rather a lot of terms 
at this order.

In the appendix we have collected the separate
contributions from both the local commutators and those involving
neighbouring vertices, for all four types of symmetrization. It 
is instructive to see what kind of terms arise in the different parts of 
the calculation and how they cancel. This may be relevant to attempts
in the continuum loop representation of quantum gravity to reproduce
similar quantum commutators. Note also that in order to derive (4.1)
below from the formulae given in the appendix, some partial integrations
had to be performed.

After a lot of algebra one finds that, independent of the symmetrization, 
the final result is given by

$$
\eqalign{
\lim_{a\rightarrow 0} 
[ \sum_{n} &\hat{{\cal V}}[N,n), \sum_{m} \hat{{\cal V}}[M,m)]|_{n_0}=\cr
a^{3}(&N^{1}(\partial_{1}M^{2}) (\hat F_{23}^{i}\hat E_{i}^{3}-
\hat F_{12}^{i}\hat E_{i}^{1})+
M^{2}(\partial_{2}N^{1}) (\hat F_{31}^{i} \hat E_{i}^{3}-
\hat F_{12}^{i} \hat E_{i}^{2})-\cr
&N^{1}M^{2}\hat F_{12}^{i}(\hat\nabla_{1}\hat E_{i}^{1}+
\hat\nabla_{2}\hat E_{i}^{2}+\hat\nabla_{3}\hat E_{i}^{3})+\cr
&N^{1}M^{2}(\hat\nabla_{1}\hat F_{23}^{i}+\hat\nabla_{2}\hat F_{31}^{i}+
\hat\nabla_{3}\hat F_{12}^{i})\hat E_{i}^{3})+O(a^{4}).}\eqno(4.1)
$$

\noindent Comparing with equation (1.4), this is the expected answer, 
without any anomalies, and up to a term proportional to the Bianchi identity.
It shows that at this level both the classical discretization and the
quantization of the diffeomorphism constraints are consistent.
The independence of the symmetrization suggests that there is a chance
that the evaluation of more complicated commutators involving also the 
Hamiltonian constraint may already yield the right result if done
in terms of the unsymmetrized lattice constraints. This is potentially
important since the calculations become even more complicated.

Let us now turn to the issue of factor-ordering.
We have already proven the absence of anomalies with momenta
ordered to the right and will now use this result to deduce what
happens for arbitrary factor-ordering. Let us adopt the notation
$({\rm Tr}\,U\tau_i)\hat p_i$ for a typical term of the quantized
diffeomorphism constraint, where $U$ denotes some plaquette loop.
We have

$$
\eqalign{
[({\rm Tr}&\,U\tau_i)\hat p_i,({\rm Tr}\,V\tau_j)\hat p_j]=\cr
&({\rm Tr}\,U\tau_i)[\hat p_i,{\rm Tr}\,V\tau_j]\hat p_j-
({\rm Tr}\,V\tau_j)[\hat p_j,{\rm Tr}\,U\tau_i]\hat p_i+
({\rm Tr}\,U\tau_i)({\rm Tr}\,V\tau_j) [\hat p_i,\hat p_j]}
\eqno(4.2)
$$

\noindent and want to investigate what happens when the momenta
are ordered to the left, i.e. whether

$$
\eqalign{
[\hat p_i&({\rm Tr}\,U\tau_i),\hat p_j({\rm Tr}\,V\tau_j)]-
\hat p_j({\rm Tr}\,U\tau_i)[\hat p_i,{\rm Tr}\,V\tau_j]+\cr
&\hat p_i({\rm Tr}\,V\tau_j)[\hat p_j,{\rm Tr}\,U\tau_i]-
[\hat p_i,\hat p_j]({\rm Tr}\,U\tau_i)({\rm Tr}\,V\tau_j) }
\eqno(4.3)
$$

\noindent vanishes or yields new terms of order $\hbar$. Rewriting
$\hat p_i({\rm Tr}\,U\tau_i)=({\rm Tr}\,U\tau_i)\hat p_i+
[\hat p_i,{\rm Tr}\,U\tau_i]$ and using equation (4.2), 
expression (4.3) becomes

$$
\eqalign{
[({\rm Tr}&\,U\tau_i)\hat p_i,[\hat p_j,{\rm Tr}\,V\tau_j]]+
[[\hat p_i,{\rm Tr}\,U\tau_i],({\rm Tr}\,V\tau_j)\hat p_j]-\cr
&[\hat p_j, ({\rm Tr}\,U\tau_i) [\hat p_i,{\rm Tr}\,V\tau_j]]-
 [\hat p_i, ({\rm Tr}\,V\tau_j) [{\rm Tr}\,U\tau_i,\hat p_j]]-
 [[\hat p_i,\hat p_j],({\rm Tr}\,U\tau_i)({\rm Tr}\,V\tau_j)]=\cr
({\rm Tr}&\,U\tau_i)[\hat p_i,[\hat p_j,{\rm Tr}\,V\tau_j]]+
 ({\rm Tr}\,V\tau_j)[[\hat p_i,{\rm Tr}\,U\tau_i],\hat p_j]-
({\rm Tr}\,U\tau_i)[\hat p_j,[\hat p_i, {\rm Tr}\,V\tau_j]]-\cr
&({\rm Tr}\,V\tau_j)[\hat p_i,[{\rm Tr}\,U\tau_i,\hat p_j]]-
[[\hat p_i,\hat p_j],{\rm Tr}\,U\tau_i]({\rm Tr}\,V\tau_j)-
({\rm Tr}\,U\tau_i)[[\hat p_i,\hat p_j],{\rm Tr}\,V\tau_j]=0}
\eqno(4.4)
$$

\noindent where we have made repeated use of the basic commutators (2.1). 
The last equality in (4.4) holds by virtue of the Jacobi identity 
satisfied by the basic quantum operators. This latter property is
not entirely trivial. For example, the holonomy and momentum
operators as they are usually defined in the continuum loop
representation do not satisfy a Jacobi identity [22,9], a feature that
gives rise to the occurrence of anomalous commutators for
certain composite operators.

It follows immediately that also for an arbitrary factor-ordering of
the lattice constraints, $\alpha ({\rm Tr}\,U\tau_i)\hat p_i +(1-\alpha )
\hat p_i ({\rm Tr}\,U\tau_i)$, $0\leq\alpha\leq 1$, no anomalies
appear. This is in particular true for the case $\alpha=\frac{1}{2}$,
where the regularized constraint operators are self-adjoint.

\vskip2cm
\line{\ch 5 Summary\hfil}

We have performed a computation of the commutator of two regularized 
diffeomorphism constraint operators in lattice gravity, and found that 
their algebra closes without anomalies in the limit of vanishing 
lattice spacing. The discretization and quantization of the classical
diffeomorphism phase space functions is not unique, but our result is 
independent of the choice of a local symmetrization and factor-ordering.
Independence of factor-ordering followed from the simple structure of the
constraints (linearity in momenta) and the fact that the basic lattice
operators satisfy the Jacobi identity. We do not expect a similar
behaviour for commutators involving also the Hamiltonian constraint.

There are not many regularization schemes for full
four-dimensional quantum gravity in which a computation of this type
could be performed. Within our lattice formulation, it would in principle
be preferable to have an exact remnant of the diffeomorphism symmetry 
at each stage of the discretization. This would enable one to study 
invariant measures and quantum states {\it before} the continuum limit is
taken. One way of proceeding is to try to identify a discrete subgroup of 
the diffeomorphism group (rather than using the discretized 
``infinitesimal" generators) for each finite lattice, an issue we are 
currently considering. Transformations of this type will presumably
be non-local in terms of the lattice variables.

\vfill\eject
%\vskip2cm
\line{\ch Appendix\hfil}

In this appendix we present some intermediate results of the commutator 
calculation, depending on the symmetrization chosen. This will be 
helpful to anybody attempting a similar calculation. We split up the
contributions into those that arise from contributions based at the 
same lattice point (the origin in our case) and those involving 
contributions based at neighbouring points on the lattice. It is
understood that $A$, $F$ and $E$ are operators.

\item{(a)} No symmetrization; contributions at the origin:

$$
\eqalign{
-a^{2}& (\, N^{1}M^{2}F_{12}^{i}(E_{i}^{1} +E_{i}^{2}+ E_{i}^{3}) +
N^{1}M^{2} ( F_{12}^{i} +F_{23}^{i} +F_{31}^{i})E_{i}^{3}\, ) -\cr
a^{3}&(\, (M^{2}\tilde N^{1} +N^{1}\tilde M^{2}) F_{12}^{i} 
(E_{i}^{1} +E_{i}^{2} + E_{i}^{3})  +
(M^{2}\tilde N^{1} +N^{1}\tilde M^{2}) (F_{12}^{i} +F_{23}^{i} 
+F_{31}^{i}) E_{i}^{3} +\cr
&\frac{1}{2}N^{1}M^{2}
 (\nabla_{1}+\nabla_{2}) F_{12}^{i}(E_{i}^{1}+E_{i}^{2}+E_{i}^{3})+\cr
&\frac{1}{2}N^{1}M^{2} ( (\nabla_{1}+\nabla_{2}) F_{12}^{i} +
(\nabla_{2}+\nabla_{3}) F_{23}^{i}+ (\nabla_{1}+\nabla_{3}) 
F_{31}^{i})E_{i}^{3}+\cr
&N^{1}M^{2}F_{12}^{i} (\nabla_{1}E_{i}^{1}+\nabla_{2}E_{i}^{2}+ 
\nabla_{3}E_{i}^{3})+
N^{1}M^{2}(F_{12}^{i}+F_{23}^{i}+F_{31}^{i}) \nabla_{3}E_{i}^{3}\,)+O(a^{4})}
\eqno(A.1)
$$

\noindent contributions involving neighbouring vertices:

$$
\eqalign{
a^{2} &(\,N^{1}M^{2}  F_{12}^{i}(E_{i}^{1} +E_{i}^{2}+ E_{i}^{3})+
N^{1}M^{2} ( F_{12}^{i} +F_{23}^{i} +F_{31}^{i})E_{i}^{3} \, ) +\cr
a^{3}&(\, (M^{2}\tilde N^{1} +N^{1}\tilde M^{2}) F_{12}^{i}
(E_{i}^{1} +E_{i}^{2} + E_{i}^{3})  +
(M^{2}\tilde N^{1} +N^{1}\tilde M^{2}) (F_{12}^{i} +F_{23}^{i} 
+F_{31}^{i}) E_{i}^{3}  +\cr
&\frac{1}{2} N^{1}M^{2} (\partial_{1}F_{12}^{i}+3 \epsilon^{i}{}_{jk} 
 A_{1}^{j} F_{12}^{k}+\nabla_{2}F_{12}^{i}) E_{i}^{1} +
\frac{1}{2} N^{1}M^{2} (\partial_{2}F_{12}^{i}+3 \epsilon^{i}{}_{jk} 
 A_{2}^{j} F_{12}^{k}+\nabla_{1}F_{12}^{i}) E_{i}^{2}+\cr
&\frac{1}{2} N^{1}M^{2}  ( 2 (\nabla_{1}+\nabla_{2}) F_{12}^{i} +
(\nabla_{2}+\nabla_{3}) F_{23}^{i}+ (\nabla_{1}+\nabla_{3}) 
F_{31}^{i})E_{i}^{3}-\cr
&N^{1}M^{2} F_{12}^{i} (\nabla_{1}E_{i}^{1}+\nabla_{2}E_{i}^{2}+ 
 \nabla_{3}E_{i}^{3})-
N^{1}M^{2} F_{12}^{i} (\partial_{1}E_{i}^{1}+\partial_{2}E_{i}^{2}+ 
 \partial_{3}E_{i}^{3})+\cr
&N^{1}M^{2} (F_{12}^{i}+F_{23}^{i}+F_{31}^{i}) \nabla_{3}E_{i}^{3}-
N^{1}M^{2}(F_{12}^{i} \partial_{3}E_{i}^{3} +F_{23}^{i} 
\partial_{1}E_{i}^{3}+ F_{31}^{i} \partial_{2}E_{i}^{3})-\cr
&N^{1}(\partial_{1}M^{2})F_{12}^{i}E_{i}^{1}-
M^{2}(\partial_{2}N^{1})F_{12}^{i}E_{i}^{2}-
N^{1}(\partial_{3}M^{2})F_{12}^{i}E_{i}^{3}-
M^{2}(\partial_{3}N^{1})F_{12}^{i}E_{i}^{3}-\cr
&M^{2}(\partial_{1}N^{1})F_{23}^{i}E_{i}^{3}-
N^{1}(\partial_{2}M^{2})F_{31}^{i}E_{i}^{3}+
N^{1}M^{2}\epsilon_{ijk}  (2 A_{3}^{j}F_{12}^{k}+ A_{2}^{j}F_{31}^{k}+
 A_{1}^{j}F_{23}^{k})E^{3i}\,)+O(a^{4})}\eqno(A.2)
$$

\item{(b)} Symmetrization over shift functions; contributions at the origin:

$$
\eqalign{
-a^{2}& (\frac{1}{2} N^{1}M^{2} F_{12}^{i}(E_{i}^{1} +E_{i}^{2}+ 2 E_{i}^{3}) 
  ) -\cr
a^{3}&( \, \frac{1}{4} (M^{2}\tilde N^{1} +M^{2}\tilde{\tilde{N}}{}^{1} +
 2N^{1}\tilde M^{2}) F_{12}^{i} E_{i}^{1} + 
\frac{1}{4} (2M^{2}\tilde N^{1} +N^{1}\tilde 
M^{2}+N^{1}\tilde{\tilde{M}}{}^{2}) F_{12}^{i} E_{i}^{2} +\cr
&\frac{1}{4} (3M^{2}\tilde N^{1} +M^{2}\tilde{\tilde{N}}{}^{1} +
       3N^{1}\tilde M^{2}+N^{1}\tilde{\tilde{M}}{}^{2}) F_{12}^{i}E_{i}^{3} +\cr
&\frac{1}{2} M^{2}(\tilde N^{1} -\tilde{\tilde{N}}{}^{1}) F_{23}^{i} E_{i}^{3} +
\frac{1}{2} N^{1}(\tilde M^{2} -\tilde{\tilde{M}}{}^{2}) 
F_{31}^{i}E_{i}^{3} +\cr
&\frac{1}{4} N^{1}M^{2} ((\nabla_{2}F_{12}^{i})E_{i}^{1}+
(\nabla_{1}F_{12}^{i})E_{i}^{2}+ 
((\nabla_{1}+\nabla_{2}) F_{12}^{i})E_{i}^{3}) +\cr
&\frac{1}{2}F_{12}^{i} (\nabla_{1}E_{i}^{1}+\nabla_{2}E_{i}^{2}+ 
2\nabla_{3}E_{i}^{3})\, )+O(a^{4})}\eqno(A.3)
$$

\noindent contributions involving neighbouring vertices:

$$
\eqalign{
a^{2}& (\frac{1}{2} N^{1}M^{2} F_{12}^{i}(E_{i}^{1} +E_{i}^{2}+ 2 E_{i}^{3}) 
  ) +\cr
a^{3}&( \, \frac{1}{4} (M^{2}\tilde N^{1} +M^{2}\tilde{\tilde{N}}{}^{1} +
 2N^{1}\tilde M^{2}) F_{12}^{i} E_{i}^{1} + 
\frac{1}{4} (2M^{2}\tilde N^{1} +N^{1}\tilde 
M^{2}+N^{1}\tilde{\tilde{M}}{}^{2}) F_{12}^{i} E_{i}^{2} +\cr
&\frac{1}{4} (3M^{2}\tilde N^{1} +M^{2}\tilde{\tilde{N}}{}^{1} +
       3N^{1}\tilde M^{2}+N^{1}\tilde{\tilde{M}}{}^{2}) F_{12}^{i} E_{i}^{3}+\cr
&\frac{1}{2} M^{2}(\tilde N^{1} -\tilde{\tilde{N}}{}^{1}) F_{23}^{i} E_{i}^{3}+
\frac{1}{2} N^{1}(\tilde M^{2} -\tilde{\tilde{M}}{}^{2}) 
F_{31}^{i} E_{i}^{3}+\cr
&\frac{1}{4} N^{1}M^{2} ((\nabla_{2}F_{12}^{i})E_{i}^{1}+
(\nabla_{1}F_{12}^{i})E_{i}^{2}+ 
((\nabla_{1}+\nabla_{2}) F_{12}^{i})E_{i}^{3})+\cr
&N^{1}M^{2}\epsilon_{ijk} (A_{1}^{j}F_{12}^{k}E^{1i}+
 A_{2}^{j}F_{12}^{k}E^{2i}) +
\frac{1}{2}F_{12}^{i} (\nabla_{1}E_{i}^{1}+\nabla_{2}E_{i}^{2}+ 
2\nabla_{3}E_{i}^{3})-\cr
&N^{1}M^{2} F_{12}^{i} (\partial_{1}E_{i}^{1}+\partial_{2}E_{i}^{2}+ 
 \partial_{3}E_{i}^{3})-
N^{1}M^{2}(F_{12}^{i} \partial_{3}E_{i}^{3} +F_{23}^{i} 
\partial_{1}E_{i}^{3}+ F_{31}^{i} \partial_{2}E_{i}^{3})-\cr
&N^{1}(\partial_{1}M^{2})F_{12}^{i}E_{i}^{1}-
M^{2}(\partial_{2}N^{1})F_{12}^{i}E_{i}^{2}-
N^{1}(\partial_{3}M^{2})F_{12}^{i}E_{i}^{3}-
M^{2}(\partial_{3}N^{1})F_{12}^{i}E_{i}^{3}-\cr
&M^{2}(\partial_{1}N^{1})F_{23}^{i}E_{i}^{3}-
N^{1}(\partial_{2}M^{2})F_{31}^{i}E_{i}^{3}+
N^{1}M^{2}\epsilon_{ijk} (2 A_{3}^{j}F_{12}^{k}+ A_{2}^{j}F_{31}^{k}+
 A_{1}^{j}F_{23}^{k})E^{3i} \,)+O(a^{4})}\eqno(A.4)
$$

\item{(c)} Symmetrization over momenta; contributions at the origin:

$$
\eqalign{
&a^{2} (\, -\frac{1}{2}N^{1}M^{2} E_{i}^{3} ( F_{23}^{i} +F_{31}^{i})\, ) +
a^{3}(\, -\frac{1}{2} (M^{2}\tilde N^{1} +N^{1}\tilde M^{2})
  E_{i}^{3} (F_{23}^{i} +F_{31}^{i}) -\cr
&\frac{1}{4}N^{1}M^{2}E_{i}^{3} (\nabla_{2} F_{23}^{i}+ 
  \nabla_{1} F_{31}^{i} )-
\frac{1}{2}F_{12}^{i} (\nabla_{1}E_{i}^{1}+\nabla_{2}E_{i}^{2}+ 
2\nabla_{3}E_{i}^{3})\, )+O(a^{4})}\eqno(A.5)
$$

\noindent contributions involving neighbouring vertices:

$$
\eqalign{
a^{2} &(\, \frac{1}{2}N^{1}M^{2} ( F_{23}^{i} +F_{31}^{i}) E_{i}^{3}\, ) +\cr
a^{3} &(\, \frac{1}{2} (M^{2}\tilde N^{1} +N^{1}\tilde M^{2})
   (F_{23}^{i} +F_{31}^{i})E_{i}^{3} +
N^{1}M^{2}\epsilon_{ijk} (A_{1}^{j}E^{1i}+ A_{2}^{j})F_{12}^{k}E^{2i}) +\cr 
&\frac{1}{4}N^{1}M^{2}(\nabla_{2} F_{23}^{i}+ 
  \nabla_{1} F_{31}^{i} )E_{i}^{3} +
\frac{1}{2}F_{12}^{i} (\nabla_{1}E_{i}^{1}+\nabla_{2}E_{i}^{2}+ 
2\nabla_{3}E_{i}^{3})-\cr
&N^{1}M^{2} F_{12}^{i} (\partial_{1}E_{i}^{1}+\partial_{2}E_{i}^{2}+ 
 \partial_{3}E_{i}^{3})-
N^{1}M^{2}(F_{12}^{i} \partial_{3}E_{i}^{3} +F_{23}^{i} 
\partial_{1}E_{i}^{3}+ F_{31}^{i} \partial_{2}E_{i}^{3})-\cr
&N^{1}(\partial_{1}M^{2})F_{12}^{i}E_{i}^{1}-
M^{2}(\partial_{2}N^{1})F_{12}^{i}E_{i}^{2}-
N^{1}(\partial_{3}M^{2})F_{12}^{i}E_{i}^{3}-
M^{2}(\partial_{3}N^{1})F_{12}^{i}E_{i}^{3}-\cr
&M^{2}(\partial_{1}N^{1})F_{23}^{i}E_{i}^{3}-
N^{1}(\partial_{2}M^{2})F_{31}^{i}E_{i}^{3}+
N^{1}M^{2}\epsilon_{ijk} (2 A_{3}^{j}F_{12}^{k}+ A_{2}^{j}F_{31}^{k}+
 A_{1}^{j}F_{23}^{k}) E^{3i}\,)+O(a^{4})}\eqno(A.6)
$$

\item{(d)} Symmetrization over both shift functions and momenta; 
contributions at the origin:
  
$$
\eqalign{
-a^{3}&(\, 
\frac{1}{4} M^{2}(\tilde N^{1} -\tilde{\tilde{N}}{}^{1}) F_{23}^{i} E_{i}^{3}+
\frac{1}{4} N^{1}(\tilde M^{2} -\tilde{\tilde{M}}{}^{2}) 
F_{31}^{i} E_{i}^{3}+\cr
&\frac{1}{2}F_{12}^{i} (\nabla_{1}E_{i}^{1}+\nabla_{2}E_{i}^{2}+ 
2\nabla_{3}E_{i}^{3})\, )+O(a^{4})}\eqno(A.7)
$$

\noindent contributions involving neighbouring vertices:

$$
\eqalign{
a^{3}&(\, 
\frac{1}{4} M^{2}(\tilde N^{1} -\tilde{\tilde{N}}{}^{1}) F_{23}^{i} E_{i}^{3}+
\frac{1}{4} N^{1}(\tilde M^{2} -\tilde{\tilde{M}}{}^{2})
F_{31}^{i}  E_{i}^{3} +\cr
&N^{1}M^{2}\epsilon_{ijk} (A_{1}^{j}E^{1i}+ A_{2}^{j})F_{12}^{k}E^{2i} + 
\frac{1}{2}F_{12}^{i} (\nabla_{1}E_{i}^{1}+\nabla_{2}E_{i}^{2}+ 
2\nabla_{3}E_{i}^{3})-\cr
&N^{1}M^{2} F_{12}^{i} (\partial_{1}E_{i}^{1}+\partial_{2}E_{i}^{2}+ 
 \partial_{3}E_{i}^{3})-
N^{1}M^{2}(F_{12}^{i} \partial_{3}E_{i}^{3} +F_{23}^{i} 
\partial_{1}E_{i}^{3}+ F_{31}^{i} \partial_{2}E_{i}^{3})-\cr
&N^{1}(\partial_{1}M^{2})F_{12}^{i}E_{i}^{1}-
M^{2}(\partial_{2}N^{1})F_{12}^{i}E_{i}^{2}-
N^{1}(\partial_{3}M^{2})F_{12}^{i}E_{i}^{3}-
M^{2}(\partial_{3}N^{1})F_{12}^{i}E_{i}^{3}-\cr
&M^{2}(\partial_{1}N^{1})F_{23}^{i}E_{i}^{3}-
N^{1}(\partial_{2}M^{2})F_{31}^{i}E_{i}^{3}+
N^{1}M^{2}\epsilon_{ijk} (2 A_{3}^{j}F_{12}^{k}+ A_{2}^{j}F_{31}^{k}+
 A_{1}^{j}F_{23}^{k})E^{3i} \,)+O(a^{4})}\eqno(A.8)
$$

The more one symmetrizes, the more cancellations occur already individually 
among terms based at the origin and among terms coming from the more non-local
commutators. For example, terms of order
$a^{2}$ still appear in cases (a), (b) and (c), but no longer when the
totally symmetrized lattice operators are used. This is in line with the 
general rule
that for lattice discretizations the convergence properties in the
continuum limit are improved when one symmetrizes the lattice
expressions over positive and negative lattice directions. Note also
that terms containing partial derivatives of the shift functions
$M$ and $N$ (that are crucial in obtaining the correct commutator) 
come from the commutators involving neighbouring sites.

\vskip2cm

\vfill\eject

\line{\ch References\hfil}
\vskip1cm

\item{[1]} N.C. Tsamis and R.P. Woodard: The factor-ordering problem
  must be regulated, {\it Phys. Rev.} D36 (1987) 3641-50.
  
\item{[2]} C. Rovelli and L. Smolin: Loop space representation of
  quantum general relativity, {\it Nucl. Phys.} B331 (1990) 80-152.

\item{[3]} A. Ashtekar and C.J. Isham: Representations of the holonomy
  algebras of gravity and non-Abelian gauge theories, 
  {\it Class. Quant. Grav.} 9 (1992) 1433-67.

\item{[4]} A. Ashtekar {\it et al.}: Quantization of diffeomorphism invariant 
  theories, {\it J. Math. Phys.} 36 (1995) 6456-93. 
 
\item{[5]} T. Thiemann: Anomaly-free formulation of nonperturbative 
  four-dimensional Lorentzian quantum gravity, {\it Phys. Lett.} B380
  (1996) 257-64.

\item{[6]} L. Smolin: The classical limit and the form of the 
  hamiltonian constraint in nonperturbative quantum gravity,
  Penn State U. {\it preprint} CGPG-96/9-4.

\item{[7]} R. Loll: talk given at the Workshop on Quantum Gravity,
  Banach Centre, Warsaw, June 1997.

\item{[8]} D. Marolf: talk given at the Workshop on Quantum Gravity,
  Banach Centre, Warsaw, June 1997.
   
\item{[9]} A. Ashtekar, A. Corichi, J. Lewandowski and J. Zapata,
  to appear.

\item{[10]} R. Loll: Further results on geometric operators in quantum
  gravity, {\it Class. Quant. Grav.} 14 (1997) 1725-41.

\item{[11]} P. Renteln: Some results of SU(2) spinorial lattice
  gravity, {\it Class. Quant. Grav.} 7 (1990) 493-502.
  
\item{[12]} P. Renteln and L. Smolin: A lattice approach to spinorial
  quantum gravity, {\it Class. Quant. Grav.} 6 (1989) 275-94.

\item{[13]} O. Bostr\"om, M. Miller and L. Smolin: A new discretization of
  classical and quantum general relativity, Syracuse U. {\it preprint} 
  SU-GP-93-4-1.
  
\item{[14]} R. Loll: Non-perturbative solutions for lattice quantum gravity,
  {\it Nucl. Phys.} B444 (1995) 619-39.

\item{[15]} K. Ezawa: Multi-plaquette solutions for discretized 
  Ashtekar gravity, {\it Mod. Phys. Lett.} A11 (1996) 2921-32.
  
\item{[16]} R. Loll: A real alternative to quantum gravity in loop
  space, {\it Phys. Rev.} D54 (1996) 5381-4.

\item{[17]} H. Fort, R. Gambini and J. Pullin: Lattice knot theory and quantum
  gravity in the loop representation, {\it Phys. Rev.} D56 (1997) 2127-43.

\item{[18]} J.F. Barbero G.: Real Ashtekar variables for Lorentzian
  signature space-times, {\it Phys. Rev.} D51 (1995) 5507-10.

\item{[19]} A. Ashtekar: New variables for classical and quantum
  gravity, {\it Phys. Rev. Lett.} 57 (1986) 2244-7; A new 
  Hamiltonian formulation of general relativity, {\it Phys.
  Rev.} D36 (1987) 1587-1603.

\item{[20]} D. Giulini: Ashtekar variables in classical general 
  relativity, in {\it Canonical gravity: from classical to quantum},
  eds. J. Ehlers and H. Friedrich (Springer, Berlin, 1994).  

\item{[21]} J. Kogut and L. Susskind: Hamiltonian formulation of
  Wilson's lattice gauge theories, {\it Phys. Rev.} D11 (1975)
  395-408; J.B. Kogut: The lattice gauge theory approach
  to quantum chromodynamics, {\it Rev. Mod. Phys.} 55 (1983) 775-836.

\item{[22]} A. Ashtekar: talk given at the ESI workshop, Vienna,
  July 1996.

\end